\documentstyle[preprint,aps]{revtex}
\draft
\begin{document}
\title{Effect of the Electromagnetic Environment on Arrays of Small Normal Metal Tunnel Junctions: 
Numerical and Experimental Investigation}

\author{Sh. Farhangfar,  A. J. Manninen and J. P. Pekola}

\address{Department of Physics, University of Jyv\"{a}skyl\"{a}, P.O. Box 35 (Y5), FIN-40351 
Jyv\"{a}skyl\"{a}, Finland}

\date{\today}

\maketitle

\begin{abstract}
We present results of a set of experiments to investigate the effect of dissipative external electromagnetic 
environment on tunneling in linear arrays of junctions in the weak tunneling regime. The influence of this resistance 
decreases as the number of junctions in the chain increases and ultimately becomes negligible. Further, there is 
a value of external impedance, typically $\sim 0.5$ k$\Omega$, at which the half-width of the zero-voltage dip in 
the conductance curve shows a maximum. Some new analytical formulae, based on the phase-correlation theory, 
along with numerical results will be presented.
\end{abstract}

\vskip1.5cm
\newpage 

In recent years large attention has been paid to the role of electromagnetic environment on 
charging effects in small tunnel junctions, both theoretically and experimentally \cite{alt91}-\cite{wahlgren}. 
Yet, arrays of such tunnel 
junctions with well-defined external impedances have not been extensively discussed. This is partly because the   
theoretical formulation of such arrays is more elaborate: there are, e.g., cotunneling effects, inhomogeneities, and 
background charges, which along with the effects of environment make such an investigation rather difficult in 
general terms. This lack 
of theoretical predictions, in turn, has decelerated  experimental  
search for observation of new features in arrays. In this letter we will attempt to fill part of this gap by 
demonstrating a set of experimental observations and a comparison of them to the results obtained from the 
already existing phase-correlation (PC) theory for single tunnel junctions, which we have 
now extended to analyze junction arrays numerically. This analysis is important in setting limits to the systematic 
error of the reading of the Coulomb blockade primary thermometer \cite{pek94}.    

According to the PC theory, the tunneling rate through the {\it k}th junction of a completely symmetric array 
with $C_{k}\equiv C$, $ R_{T,k}\equiv R_{T}$ and $C_{0,k}=0$ \cite{Cstray} (see Fig. 1), in the weak-
tunneling regime, ${R_{T, k}}\gg { {R_K}\equiv{h/e^2} },$ can be written as a convolution integral of the 
form \cite{c2}

\begin{equation}\label{1}
{\Gamma^ {\pm}_k}(\{n\}) \equiv  {\Gamma}  ({\delta F^{\pm}_{k}}, \{ n \}) =
 { \frac   {1} { { e^2} R_{T} }  } 
{\int_{-\infty}^{+\infty}}{dE} \frac{E}{1-e^{-\beta E}}
P_{k}(-{\delta F_k^\pm}-E).
\end{equation}
Here ${\delta} F_k^{{+\scriptstyle\smallskip\atop(-)}}$ is the change in free energy of the array when an 
electron tunnels to right (left),  $\{ n \} \equiv \{ {n_1}, {n_2}, \ldots,{n_{N-1}}\}$ designates 
the charge configuration on the islands, and
$P_{k}(E)\equiv{(2\pi \hbar)}^{-1}\int_{-\infty}^{+\infty}dt \mbox {e}^{J_{k}(t)+i\frac{E}{\hbar}t}$ 
is the probability density for the electron to exchange energy $E$ with the environment. 
The correlation function $J_{k}(t)$, which accounts for the environment of the {\it k}th junction, is given by 
\begin{equation}\label{2}
J_{k}(t)= 2\int_{-\infty}^{\infty}
\frac{d\omega}{\omega}
\frac{\mbox {Re} [Z_t^k(\omega)]}{R_K}
\frac{\mbox{e}^{-i\omega t}-1}{1-\mbox{e}^{-\beta\hbar\omega}},
\end{equation}
where $Z_{t}^k(\omega)$ is the total impedance of the circuit as seen by this junction, and $\beta\equiv 
({k_B}T)^{-1}$. 

By applying Fourier transform techniques to Eq. (1) one obtains 
\vskip .45 cm
\begin{equation} \label {3}
{\Gamma^{\pm}_k}(\{n\})   =
{   \frac  {1}{{e^2}R_{T}}  }
\{
    {\frac {1}{\beta}}+{\frac{ {-\delta F_k^\pm}-{{i\hbar J_k^{\prime}}(0)} } {2} }-
    {\frac {\pi}{2{\beta^2}\hbar}}
     {\int_{-\infty}^{+\infty} }dt {   \frac  {e^{[J_{k}(t)-i{\delta F_k^\pm}{\frac{t}{\hbar}}]} - 1}      
                                                                       { { \sinh^2} ({ \frac{\pi t}{\beta \hbar} })    }      }
\}.
\end{equation}
This equation is central in the following numerical calculations. With $\Gamma_k$'s 
and the algorithm in \cite{hirvi96} with ${R_\Sigma}\equiv {\sum_{k=1}^N}{R_{T,k}}$, 
we find the current $I$ through the array in equilibrium as
\vskip .55cm
\begin{equation}\label{4}
I =
\sum_{{\{n\}_{visited}}} \{
e\frac
{\sum_{k=1}^{N}[{\Gamma_k^+}( \{n\}) - {\Gamma_k^-}( \{n\}) ]. {  \frac{R_{T,k}}{R_\Sigma}  }    }
{ \sum_{k=1}^{N}[{\Gamma_k^+}( \{n\}) + {\Gamma_k^-}( \{n\}) ]     }
\}
\frac{1}{     \sum \limits_{\scriptstyle{{\{n\}_{visited}}}} 
   {( \sum_{k=1}^{N}[{\Gamma_k^+}( \{n\}) + {\Gamma_k^-}( \{n\}) ] )}^{-1}    }. 
\end{equation}
\vskip .4cm
In the expression above, starting initially from an arbitrary configuration, the states visited, 
$\{n\}_{visited}$, over which the outer sums run through,  
are obtained by  dividing the interval $[0,1]$ 
into segments proportional to $\Gamma^{\pm}_k$'s in each current state,
and drawing a random number $r$ in the interval. For sequential tunneling, 
the segment to which $r$ corresponds to will specify the junction through which the tunneling event happens and 
the tunneling direction. This way the distribution of $\{n\}_{visited}$ will be statistically collected, and it will allow 
one to calculate the sums (now weighted by the distribution) according to Eq. (\ref{4}) similarly to what has been 
done in \cite{hirvi96}. 

In the case of a symmetric two-junction array we will  also use a simpler algorithm, 
described in \cite{hirvi96}, to obtain  the probability of  finding $n$ excess electrons on the island, 
$\sigma(\{n\})$, and 
finally the equilibrium current $I$ through the array
\begin{equation}\label{5}
 I={I_k}= e{\sum_{n=-\infty}^{\infty}\sigma(\{n\})[{\Gamma_k^+}( \{n\}) - {\Gamma_k^-}( \{n\}) ]}.
\end{equation}

Assuming a completely symmetric array and a purely resistive environment $R_e$, the common real part of the 
total impedance, $Z_{t}(\omega)\equiv Z_{t}^k(\omega)$, in Eq. (2) reduces to the simple form 
${\mbox{Re}}[Z_{t}(\omega)]= { {R_e}/ [{{\displaystyle({\omega}/{\omega_c})^{2}} +
{{N^2}}}]  }$,
with ${\omega_c}\equiv 1/{R_e}C$. It is already suggested by this equation that the effect of the environment 
decreases with increasing $N$ and becomes vanishingly small for long arrays. Furthermore, using the partial 
sum expansion of $\coth(x)$ (or by applying Cauchy's integral theorem), 
one can evaluate ${J}(t)\equiv{J_k}(t)$ in Eq. (2) as

\begin{equation}\label{7}  
{J}(t)=\frac{\pi}{N^2} \frac{R_e}{R_K} \{ (1-e^{-\mid{N}{\omega_c}t\mid })
[\cot(\frac{\beta \hbar N\omega_c}{2}) - i] - 
\frac{2{\mid}t\mid}{\beta\hbar}+
4\sum_{n=1}^{\infty} \frac { {(N\omega_c)}^2  (1-e^{-\mid{N}{\omega_n}t\mid }) }
 { 2\pi n [{\omega_n}^2 -{(N{\omega_c}})^2 ]}\},
\end{equation}
where ${\omega_n}\equiv {2\pi n/\beta \hbar}$ are Matsubara frequencies. The above equation is a  
straightforward extension of the result obtained for a single tunnel junction in resistive environment 
\cite {joyez,grabert}. In our numerical calculations we have used the generally valid formula in Eq. (2), 
but equivalence of results is checked both by direct comparison of the numerical values derived from 
Eqs. (2) and (6), and by comparing the final conductance curves; the results from the two methods are 
indistinguishable.

Figure 1 shows a schematic view of an array with its bias circuitry. To check for the consistency of results for 
arrays with different number of junctions, each sample included 
one {\sl pair} of arrays with about 3 $\mu$m space in between. Each pair had a different 
number of junctions, typically $N=1$, 2, $N=2$, 8, and $N=2$, 20 (only samples with $N=2$, 8 and 
$N=2$, 20 are shown in Table I). The array consisted of ${\rm A l/AlO_{x}/Al}$ tunnel junctions ($0.01-0.05$ 
$\mu$m$^2$) with four chromium resistors, $Z_{e,j}(\omega)={R_j}$, at a distance of 2 $\mu$m, two at each 
end of the array. Cr resistances of $1-20$ k$\Omega$ can be easily obtained by adjusting the width ($\sim 100$ 
nm), length ($\sim 2$ $\mu$m), and thickness ($\sim$ $3-8$ nm) of chromium films. The equivalent environment 
resistance ({\it additional to} the natural free space like impedance of $\sim 100$ $\Omega$) will,  then, be 
$R_e=R_1R_2/(R_1+R_2) + R_3R_4/(R_3+R_4)$. Samples were made by e-beam lithography and three-angle 
evaporation techniques. All measurements were carried out at 4.2 K. More details of measurement techniques 
are presented in \cite{farhangfar}. 

Figure 2 shows measured data for a typical sample with $N=2$, ${R_T}=44.9$ k$\Omega$, ${R_e}=1.12$ 
k$\Omega$ and $C=2.25$ fF (sample 9A in Table I (a)). $C$ was determined by searching 
for the best fit of the measured depth of the zero-bias anomaly to the value given by Eqs. ($1-3$); those 
equations 
are in agreement with the zero-bias results of \cite{joyez}. Each point in the simulated $IV$-curve (not shown in 
the figure) was obtained as a result of $1000$ (pseudo)random tunneling events. Conductance curve (CC) was 
obtained by numerical derivation of the $IV$-curve. It is worth mentioning that even with much smaller number of 
draws, e.g. $10$, the result agrees within better than $1.5\%$ accuracy (with respect to the height and width of 
the CC dip) with those obtained from ''long simulations''. The time needed for each step, with a reasonable 
partition of voltage interval and using a regular desktop computer, was about $1$ minute. The result with $N=2$ 
obtained from Eq. ($5$), based on the algorithm presented in \cite{hirvi96}, is identical with 
that of a more comprehensive method described above. 

In Fig. 3 we have drawn the normalized half-width of the zero-bias minimum of the conductance curve, 
$V_{1/2}$, as a function of the environment resistance for different pairs of samples. By normalization we mean 
scaling the half-widths by those derived for arrays without an external impedance, i.e., scaling by 
$V_{{1/2},0}\equiv {5.439N{k_B}T/e}$ \cite{pek94}. Usually, we have done measurements for each 
sample in six different combinations of current and voltage probes (four-probe measurement) and there are 
variations between the different combinations shown by the error bars. The unequal height of the error bars for 
different samples is due to this. For $N=2$, $V_{1/2}$ shows a sharp maximum at around $R_e \simeq 1$ 
k$\Omega$. For longer arrays, $N=8$ and $N=20$, the dependence is much weaker, and $V_{1/2}$ 
stays close to $V_{1/2,0}$, which is directly demonstrating the advantage of using long arrays in Coulomb 
blockade thermometry. In each case (unambiquously only for $N=2$, though), the normalized $V_{1/2}$ 
approaches unity with $R_e \rightarrow 0$ and for $R_e \rightarrow \infty$. This is in agreement with 
predictions of the PC theory. We will discuss this in further detail below. 

In the same figure the results obtained by numerical simulation, and for $N=2$ by the direct calculation described 
above, are depicted. In spite of the overall agreement between experiment and theory, there is a noticeable 
discrepancy between the measured and the predicted value of $V_{1/2, max}$, the widest half-width, of the two-
junction array. In this case ($N=2$) we could calculate the half-width with the two methods described above, 
without a noticeable difference between them. For $N=20$ the absolute difference between experiment and 
theory is smaller; the predicted peak itself is very small, less than $1\%$ (see inset of Fig. 3). Comparison of 
experimental data to those derived from the theory for $N=2$ shows that the former produce even $15\%$ wider 
conductance curves (Fig. 3). Cotunneling and other higher order tunneling effects may play some role in our 
samples, because the junction resistances are not large as compared to $R_K$. While in the 
simpler case of a two-junction array without electromagnetic environment higher order tunneling tends to broaden 
the half-width of the CC dip \cite{k}, making such comparison between our data and the theoretical study of 
cotunneling effects in an array with dissipative environment \cite{odin2} is beyond the scope of this work and is 
not done here. There is also a difference in the shape of the ''shoulders'' of the measured conductance curves for 
$N=2$ in particular (see Fig. 2) as compared to the theory. Such a distortion of shape can be caused by the 
nonuniform size distribution of junctions in the array, which we have studied in the case of no external impedance 
\cite {pek94,hirvi95}. Experimentally the intercomparison of the depths in different samples is difficult because 
the size of the junctions varies from sample to sample, and this gives the main contribution to the depth variation.

Next, let us consider the conductance curve in more detail. Using the time-domain formulation of 
single electron tunneling presented in \cite{joyez,odin} we repeated calculation of the high-temperature 
conductance $G(V)$ of the symmetric $N$-junction array without stray capacitance in the limit of large $R_e$ 
($Nu_{N}\ll{R_e}/{R_K}$; $ u_{N} \equiv [(N-1)/N][{e^2}/Ck_{B}T ]$). The result reads:
$ \frac{G(V)}{G_T}= {1-u_{N}g(v)}$
with ${G_T}\equiv 1/N{R_T}$, $v\equiv eV\beta/N$ and $g(x)\equiv[x\sinh x -4{\sinh^{2} (x/2)}]{\hskip 
.75 mm}/{\hskip .5mm} 8{\sinh^{4}(x/2)}$. Comparison of the above expression with the corresponding one 
derived for the 
perfectly conducting environment (${R_e}=0$) in \cite{pek94}, 
${\frac{G(V)}{G_T}}= {1-\frac{N-1}{N} u_{N}g(v)}$, indicates that 
${{(\frac{\Delta G}{G_T})}_{{R_e}\to \infty}} = {\frac{N}{N-1}}
{ {(\frac{\Delta G}{G_T})}_{{R_e}\to 0} }$,
where $\frac{\Delta G}{G_T}\equiv 1-\frac{G(0)}{G_T}$ stands for the depth of the conductance dip. Here, 
again, the 
half-width has a universal value given by $V_{1/2}=V_{1/2, 0}$. The above equation for the asymptotic values 
of ${\Delta G}/{G_T}$ together with the last 
expression for $V_{1/2}$ confirms that the effect of a dissipative environment on conductance becomes 
increasingly suppressed by large $N$ and is most noticeable for a two-junction array.

Finally, let us have a closer look at the results above in view of Coulomb blockade thermometry (CBT) which is a 
primary (and secondary) thermometer based on single electron charging effects in arrays of tunnel junctions 
\cite{pek94}. The main parameter is the half-width of the CC dip. We conclude that the inaccuracy in 
temperature measurements arising from environmental effects can be made small by increasing the number of 
junctions, $N$. Our 
numerical simulations along with the experimental results depicted above show that for $N=20$ 
the temperature determined by CBT is very close to the thermodynamic temperature. In practice, in a sample 
with no intentional $R_e$ the effective value of the external impedance is of the order of free space impedance 
$R_{e, eff}< Z_{0}\simeq 377$ $\Omega$, and therefore the agreement is better than $\pm0.5\%$ (see the 
experimental data point marked by an open circle in Fig. 2).

In summary, we have studied the effect of the resistive electromagnetic environment on transport in arrays of 
normal metal tunnel junctions within the high temperature and the weak tunneling regime. Special attention has 
been paid to the half-width of the conductance curve. Overall agreement between numerical results based on the 
extension of the PC theory and experimental data has been observed. We cannot explain the quantitative 
discrepancy between theory and experiment for the strong enhancement of the width at intermediate values of 
$R_e$ in the $N=2$ case: higher order tunnelling effects have not, however, been included in the present theory. 
As a practical conclusion, we verify that the effect of environment is mostly emphasized in a two-junction array and 
can be made sufficiently small for thermometric applications by increasing the number of junctions in the array. 

We thank A. Korotkov and J. K{\"o}nig for discussions. We thank the National Graduate School in Materials 
Physics for support.
\newpage

{\bf TABLE CAPTION} 
{\bf Table I.} {\bf (a)} Parameters of the measured samples with $N=2$. Samples with the same capital 
letter, e.g., 2A, 8A, 9A and 10A, belong to the same chip. {\bf (b)} and {\bf (c)} Parameters of the measured 
samples with $N=8$ and $N=20$, respectively. Samples 4C and 1C, 7C and 2C, and 5B and 1B constitute 
pairs of samples, with a space of only about 3 $\mu$m between the respective arrays. Capacitances are 
obtained by fitting the theoretical conductance curves to the experimental ones. The only fitting parameter is the 
capacitance of the sample.   

\newpage 

{\bf FIGURE CAPTIONS}

{\bf Fig. 1.} An array of $N$ tunnel junctions in an electromagnetic environment together with its bias circuitry. 
For a symmetric array in a purely dissipative environment, ${Z_{e,j}(\omega)}=R_{j}$, and with negligible 
stray capacitances, $C_{0,k}=0$, the real part of the total impedance as seen by the $k$th junction, 
${\mbox Re}[{Z_t^k}(\omega)]$, reduces to the simple form presented in the text (Eq. (6)). 

{\bf Fig. 2.} The measured conductance curve for a two-junction array with ${R_e}=1.12$ k$\Omega$ 
(sample 9A). The curve under the data points is the theoretical conductance curve with $C=2.25$ fF (see text).        

{\bf Fig. 3.} Measured half-width of the conductance curve, normalized by $5.439Nk_{B}T/e$ (see text), for 
samples with $N=2$ (solid diamonds), $N=8$ (open triangles)  
and $N=20$ (open squares) as a function of {\it extra} (i.e., that {\sl in addition} to the 
unintentional space impedance of $\sim 100$ $\Omega$) external resistance. For comparison, a sample 
($N=20$) without 
any intentional on-chip impedance has been shown in the plot by an open circle. 
The lowermost solid curve is obtained by simulation for completely symmetric $20$-junction array with 
$C\equiv C_{k}=5.0$ fF. The uppermost solid curve is the result of simulation for a two-junction array with 
$C=2.2$ fF, whereas dotted and dashed lines correspond to $C=5.0$ fF and 
$C=1.4$ fF, 
respectively. The middle solid curve indicates the result of simulation for an eight-junction array with $C=5.0$ fF. 
The capacitances were obtained from the depth of the dip of the conductance curves. The inset shows 
the normalized half-width together with the depth of the conductance curve ($\Delta G/{G_T}$), obtained from 
the theory, for a 20-junction array. 


\begin{references}
\bibitem{Cstray}
For the $\sl k$th island of length $l$ and made of a material with permittivity $\varepsilon$, the stray 
capacitance, $C_{0,k}$, can be written as ${C_{0,k}}\sim{\varepsilon}l$. In our samples islands are made 
of aluminium and their lenghts are $\sim 1$ $\mu$m. This gives rise to $C_{0,k}\sim 0.5\times 10^{-17}$ F. 
This is two order of magnitudes smaller than typical capacitances of our samples ($\sim2-5$ fF). This is why we 
neglect stray capacitances now.
\end{references}
\end{document}